\documentclass[pra,letterpaper,twocolumn,aps,superscriptaddress,showpacs,longbibliography,nofootinbib]{revtex4-1}

\usepackage{amsmath,upgreek,amssymb,graphicx,subfigure,dsfont,color,hyperref,bbm}
\usepackage{soul}
\usepackage{enumitem}
\setitemize{leftmargin=*}
\usepackage{comment}

\usepackage[T1]{fontenc}
\usepackage{lmodern}

\usepackage{fancyhdr}
\usepackage{tikz}
\usepackage{tikz-3dplot}
\usepackage{pgfplots}
\usepackage[margin=0.5in]{geometry}
\usepackage{graphicx}
\usetikzlibrary{pgfplots.groupplots}

\pgfplotsset{compat=newest}

\usepackage{multirow}
\usepackage[english]{babel}
\usepackage{fullpage}
\usepackage{verbatim}
\usepackage{float}

\newcommand{\ii}{\mathrm{i}}
\newcommand{\e}{\mathrm{e}}

\usepackage{physics}

\graphicspath{{/home/victor/Escritorio/primerarticulo/Imagenes/}}

\DeclareMathAlphabet\mathbfcal{OMS}{cmsy}{b}{n}

\newcommand{\Ah}{\hat{\mathbf{A}}^{\bot}}

\newcommand{\com}{\mathrm{com}}

\makeatletter
\newsavebox{\@brx}
\newcommand{\llangle}[1][]{\savebox{\@brx}{\(\m@th{#1\langle}\)}%
  \mathopen{\copy\@brx\kern-0.5\wd\@brx\usebox{\@brx}}}
\newcommand{\rrangle}[1][]{\savebox{\@brx}{\(\m@th{#1\rangle}\)}%
  \mathclose{\copy\@brx\kern-0.5\wd\@brx\usebox{\@brx}}}
\makeatother

\newcommand{\omegaL}{\omega_L}
\newcommand{\DeltaL}{\Delta_L}

\usepackage{titlesec}

\titleformat*{\subsubsection}{\bfseries}

\usepackage{natbib}

\begin{document}

\title{\textit{Ab initio} quantum theory of mass defect and time dilation in trapped-ion optical clocks}

\author{V. J. Mart{\'{\i}}nez-Lahuerta}
\affiliation{Institute for Theoretical Physics and Institute for Gravitational Physics (Albert-Einstein-Institute), Leibniz University Hannover, Appelstrasse 2, 30167 Hannover, Germany}

\author{S. Eilers}
\affiliation{Institute for Theoretical Physics and Institute for Gravitational Physics (Albert-Einstein-Institute), Leibniz University Hannover, Appelstrasse 2, 30167 Hannover, Germany}

\author{T. E. Mehlst\"{a}ubler}
\affiliation{Physikalisch-Technische Bundesanstalt, Bundesallee 100, 38116 Braunschweig, Germany}
\affiliation{Institute for Quantum Optics, Leibniz University Hannover, Welfengarten 1, 30167 Hannover, Germany.}

\author{P. O. Schmidt}
\affiliation{Physikalisch-Technische Bundesanstalt, Bundesallee 100, 38116 Braunschweig, Germany}
\affiliation{Institute for Quantum Optics, Leibniz University Hannover, Welfengarten 1, 30167 Hannover, Germany.}

\author{K. Hammerer}
\affiliation{Institute for Theoretical Physics and Institute for Gravitational Physics (Albert-Einstein-Institute), Leibniz University Hannover, Appelstrasse 2, 30167 Hannover, Germany}

\begin{abstract}
 We derive a Hamiltonian for the external and internal dynamics of an electromagnetically bound, charged two-particle system in external electromagnetic and gravitational fields, including leading-order relativistic corrections. We apply this Hamiltonian to describe the relativistic coupling of the external and internal dynamics of cold ions in Paul traps, including the effects of micromotion, excess micromotion, and trap imperfections. This provides a systematic and fully quantum-mechanical treatment of relativistic frequency shifts in atomic clocks based on single trapped ions. Our approach reproduces well-known formulas for the second-order Doppler shift for thermal states, which were previously derived on the basis of semiclassical arguments. We complement and clarify recent discussions in the literature of the role of time dilation and mass defect in ion clocks.
\end{abstract}

\date\today

\maketitle

\section{Introduction}
Optical ion clocks~\cite{Ludlow2015_review} have shown systematic uncertainties below $10^{-18}$ \cite{Brewer2019}. This fulfills an early prediction of Dehmelt~\cite{Dehmelt1982} and achieves an important milestone on the way towards a possible redefinition of the SI second~\cite{Riehle2018}. Clocks at this level of uncertainty open the way to many applications, such as relativistic geodesy~\cite{lisdat_clock_2016,grotti_geodesy_2018,McGrew2018,denker_geodetic_2018,mehlstaubler_atomic_2018,wu_towards_2020,muller_using_2020}, tests of general relativity~\cite{takamoto_test_2020,delva_gravitational_2018,herrmann_test_2018}, and explorations of physics beyond the standard model~\cite{Safronova2018}. At the same time, systematic relativistic frequency shifts and their uncertainty play a significant and even dominant role~\cite{Berkeland1998,weyers_advances_2018,Rosenband2010,Peik2012,Keller2015,Chou1630,bothwell_resolving_2021}. This concerns in particular the special-relativistic second-order Doppler shift $-\mathbf{v}^2/2c^2$, which accounts for moving clocks ticking slower than stationary clocks, and the general-relativistic gravitational red shift~\cite{Ludlow2015_review}. Both shifts can be seen as an effect of time dilation, which occurs when the proper time measured by the clock atom along its world line is Lorentz transformed into the reference frame of the laboratory or that of another distant clock. This reasoning is entirely correct and rigorous, but implicitly assumes a semiclassical approach in which the center of mass motion of the atom is ascribed a classical world line and only its internal (electronic) degree of freedom is treated quantum mechanically.

An alternative perspective can be gained by placing not time dilation but the mass defect, i.e. the equivalence of internal (binding) energy and external (kinematic as well as gravitational) mass, at the center of reasoning~\cite{Zych2018}. Already from treatments of the Mössbauer effect~\cite{Josephson1960,DEHN1970}, it is known that the mass defect gives rise to a frequency shift of internal transitions that is equivalent to the second-order Doppler effect. In the context of ion clocks, this equivalence was pointed out recently in~\cite{Jan2019mdtdequiv}. The advantage of this perspective is that the mass defect can be represented by a simple Hamiltonian coupling between electronic and center of mass (COM) degrees of freedom (DOFs), which is treated as a relativistic perturbation to the standard Hamiltonian of non relativistic quantum optics. The basic formulas of the known relativistic corrections can then be reproduced on the grounds of this perturbed Hamiltonian, as shown in~\cite{Yudin2018, Haustein2019}. Treatments based on the mass defect, however, also led to considerations of possible new types of systematic shifts~\cite{Yudin2018}, and alleged fundamental limits on the accuracy of atomic clocks~\cite{Sinha2014}, beyond what is known from time dilation. Apart from these disparities, approaches relying on the mass defect have not yet obtained a rigorous treatment of the micromotion and excess micromotion, which are known to play an important role in relativistic frequency shifts in trapped ion clocks~\cite{Berkeland1998,Keller2015}. Moreover, while these approaches have the potential to provide a fully quantum-theoretic picture of relativistic shifts, there does not appear to be a self-contained derivation of the perturbed Hamiltonian that covers the case of a trapped ion to date.

We set out here to give a systematic derivation of the Hamiltonian for an ion in external electromagnetic and gravitational fields, including relativistic corrections involving external and internal DOFs, building on~\cite{Sonnleitner2018,Schwartz2019}. We apply this Hamiltonian to the context of an ion clock and give a rigorous quantum-mechanical derivation of the relativistic frequency shifts. Including the effects of micromotion and excess micromotion in the framework of Floquet theory~\cite{glauber1992,Leibfried2003_review,zel1967quasienergy,ritus1967shift,Sambe}, we reproduce, for the special case of thermal states of motion, the shifts known from~\cite{Berkeland1998,Keller2015}. Further frequency shifts or fundamental limitations do not arise.

The derivation of the Hamiltonian closely follows that in~\cite{Sonnleitner2018,Schwartz2019} and merely extends this work to composite systems with nonvanishing total charge. Starting from the classical Lagrangian of an electromagnetically bound two-particle system, this derivation establishes the mass defect as the only relevant relativistic correction to the standard quantum-optical Hamiltonian which couples COM and internal DOFs. The thus corrected Hamiltonian strictly refers to the laboratory frame and completely covers all relativistic frequency shifts. A further correction of time dilation is unnecessary. The quantum-mechanical description developed here also covers the effects of zero-point fluctuation as well as arbitrary other quantum states of motion. We hope that the logic developed here will prove useful also in other, more complex systems, such as multi-ion clocks or optical lattice clocks, to obtain a stringent analysis of special and general relativistic effects.

The article is organized as follows.
In Sec.~\ref{section2} we treat the derivation of the approximate relativistic Hamiltonian. Building on this, we apply the Hamiltonian in Sec.~\ref{section3} to the problem of a single ion in a Paul trap and draw conclusions regarding systematic shifts in frequency metrology. We summarize in Sec.~\ref{sec:Conclusions}

\section{Hamiltonian of a charged composite system in external electromagnetic and gravitational fields}\label{section2}

In this section we summarize the derivation of the Hamiltonian for an ion coupled to external electromagnetic and gravitational fields including first-order relativistic corrections. Special focus is put on relativistic coupling of internal (relative) and external (COM) DOFs. We adopt the model of Sonnleitner and Barnett \cite{Sonnleitner2018} for a hydrogenlike atomic ion as an electromagnetically bound two-body system composed of a core (charge $e_1$ and coordinates $\mathbf{r}_1$) and an electron ($e_2$ and $\mathbf{r}_2$). In contrast to Sonnleitner and Barnett, we allow for a nonvanishing net charge $Q=e_1+e_2\neq 0$ and take into account a nonzero gravitational field. For the latter we follow the treatment of  Schwartz and Giulini \cite{Schwartz2019}, who extended the calculation from Sonnleitner and Barnett by a weak gravitational background field described by the Eddington-Robertson parametrized post-Newtonian (PPN) metric, thus covering first-order relativistic corrections to the Minkowski metric. The notation used in this section corresponds to that of~\cite{Schwartz2019}. In the following, we only present the most important steps and results of the calculation. We present the main differences our work and the aforementioned derivations in Appendix~\ref{AppendixrelHam}. 

We note that the mass defect Hamiltonian can be derived also on other grounds, based, e.g., on effective field theory for composite systems~\cite{Anastopoulos2021,Zych2018} or on approaches to formalize time dilation within quantum theory~\cite{Khandelwal2020,Smith2020}. The account of \cite{Sonnleitner2018,Schwartz2019} that we follow here proceeds in the spirit of conventional atomic structure calculations and has the benefit of systematically providing all relevant relativistic corrections, not just the mass defect. Relativistic corrections due to spin are not covered here, however, and would require suitable extensions of treatments of composite particles based on the Dirac equation along the lines of~\cite{Pachucki2004,Pachucki2007}.

\subsection{Classical Lagrangian and quantization of a composite system}

Our starting point is the classical Lagrangian for two particles interacting with the electromagnetic field 
\begin{align}
    &L =~ \mathord{-} \sum_{i=1,2} m_i c \sqrt{-g_{\mu \nu}(\mathbf{r}_i) \dot{r}_i^\mu \dot{r}_i^\nu} \label{eq:Lagrangian1_grav}\\
    &+ \int \mathrm{d}^3 \mathbf{r} \, \sqrt{-g(\mathbf{r})} \Big(
    J^\mu(\mathbf{r}) A_{\mu}(\mathbf{r})
    - \frac{1}{4 \mu_0} F_{ \mu \nu}(\mathbf{r}) F^{\mu \nu}(\mathbf{r})\Big).\notag
\end{align}
Here the first line gives the Lagrangian for the point particles and the terms in the second line represent the interaction of the electromagnetic field and the particles and the Lagrangian of the electromagnetic field, respectively, which is obtained by minimally coupling the special-relativistic action for electromagnetism to a general space-time metric~\cite{Schwartz2019}. We use four-vector notation, with \(g_{\mu \nu}(\mathbf{r})\) being the metric tensor, \(g(\mathbf{r})\) its determinant, \(J^\mu(\mathbf{r})\) the electric four-current, $A_{\mu}(\mathbf{r})$ the electromagnetic four-potential, and \(F_{\mu \nu}(\mathbf{r})\) the field strength tensor. Bold symbols denote three-vectors. The Eddington-Robertson metric is defined with respect to the Minkowski metric as the background structure and has the form \cite{Schwartz2019}
\begin{align}\label{eq:metric}
\begin{split}
    g_{00}(\mathbf{r})&=-1 - 2 \frac{\phi(\mathbf{r})}{c^2} - 2 \beta \frac{\phi^2(\mathbf{r})}{c^4} + \mathcal{O}(c^{-6}), \\ 
    g_{jj}(\mathbf{r})&=1 - 2 \gamma \frac{\phi(\mathbf{r})}{c^2} + \mathcal{O}(c^{-4}) \quad (j=1,2,3).
\end{split}
\end{align}
All off-diagonal components vanish for a flat background metric up to $\mathcal{O}(c^{-5})$. The scalar Newtonian potential is denoted by \(\phi(\mathbf{r})\). We include here also the Eddington-Robertson parameters \(\beta\) and \(\gamma\) which account for possible
deviations from general relativity and fulfill $\beta=\gamma=1$ in general relativity. 

Following \cite{Schwartz2019}, the Lagrangian~\eqref{eq:Lagrangian1_grav} is written in the Coulomb gauge and expanded in inverse powers of $c$, maintaining terms up to second order. The corresponding classical Hamiltonian function is quantized canonically, which yields the approximately relativistic Hamiltonian operator for two charged particles minimally coupled to the electromagnetic field 
\begin{align}
    \hat{H}&=\sum_{i=1,2}\left( \frac{\hat{\bar{\mathbf{p}}}_i^2}{2m_i}
    +m_i \phi(\hat{\mathbf{r}}_i)+e_i\Phi\left( \hat{\mathbf{r}}_i\right)
    \right)
    +\frac{e_1e_2}{4\pi\varepsilon_0} \frac{1}{\hat{r}} \notag\\
    &-\frac{e_1e_2}{16\pi\varepsilon_0c^2m_1m_2}\left(\hat{\bar{\mathbf{p}}}_1\cdot\frac{1}{\hat{r}}\hat{\bar{\mathbf{p}}}_2+ \left(\hat{\bar{\mathbf{p}}}_1\cdot\hat{\mathbf{r}}\right)\frac{1}{\hat{r}^3} \left(\hat{\mathbf{r}}\cdot\hat{\bar{\mathbf{p}}}_2\right) \right. \notag\\
    &\quad\quad + (1\leftrightarrow 2) \bigg)\notag\\
    &+ \sum_{i=1,2} \bigg(\!-\!\frac{\hat{\bar{\mathbf{p}}}_i^4}{8m_i^3c^2}
    +\frac{2 \gamma + 1}{2 m_i c^2} \hat{\bar{\mathbf{p}}}_i \cdot \phi(\hat{\mathbf{r}}_i) \hat{\bar{\mathbf{p}}}_i \notag\\
    &\quad\quad + (2 \beta - 1) \frac{m_i \phi^2(\hat{\mathbf{r}}_i)}{2 c^2} + (\gamma + 1) \phi(\hat{\mathbf{r}}_i) \frac{e_1 e_2}{8 \pi \varepsilon_0 c^2 \hat{r}}\bigg).\label{eq:TwoBodyHamiltonian}
\end{align}
We define $\hat{\bar{\mathbf{p}}}_i=\hat{\mathbf{p}}_i-e_i\hat{\mathbf{A}}^\bot\left( \hat{\mathbf{r}}_i\right) $, $\hat{\mathbf{r}}= \hat{\mathbf{r}}_1 - \hat{\mathbf{r}}_2$, and $\hat{r} = \left| \hat{\mathbf{r}} \right| $.  The electromagnetic three-potential $\hat{\mathbf{A}}^\bot\left( \hat{\mathbf{r}}\right)$ (transverse in the Coulomb gauge) and the electric potential $\Phi\left( \hat{\mathbf{r}}\right)$ are taken as classical variables describing externally applied fields~\footnote{This has to be understood in the sense of a mean-field treatment with respect to an externally applied electromagnetic field. The effects of the radiation reaction such as spontaneous emission or Lamb shifts are lost in this approximation.}. We refer the reader to~\cite{Schwartz2019} for a comprehensive derivation of this result. The first line in Eq.~\eqref{eq:TwoBodyHamiltonian} is the nonrelativistic Hamiltonian; the following lines give the dominant relativistic corrections.

\subsection{Hamiltonian for internal and external DOF}

The Hamiltonian~\eqref{eq:TwoBodyHamiltonian} is now subjected to a Power-Zienau-Woolley transformation to change to the multipolar representation of the light-particle interaction. Without relativistic corrections, the resulting Hamiltonian can be separated into terms referring to internal and external DOFs corresponding to COM and relative coordinates. However, with relativistic corrections, these coordinates no longer separate the two DOF fully, so after this change \textit{relativistic} variants of these coordinates have to be introduced.
Since a system's energy content is part of its inertia, the proper way to discuss this problem should be to express the Hamiltonian in the center of energy frame, but this is not a canonical transformation, as is explained in \cite{Sonnleitner2018}.
Nevertheless, there exists a choice of coordinates via a canonical transformation that allows separation of COM and relative dynamics up to our order of approximation, as shown by Close and Osborn \cite{Close1970} for the case of neutral systems $Q=0$. We can show that this transformation can be generalized to the case $Q\neq 0$(see Appendix~\ref{AppendixrelHam}), and these will be the coordinates that we will use from now on.

In terms of these coordinates the Hamiltonian can be written as 
\begin{equation}\label{eq:separatedH}
	\hat{H} = \hat{H}_{\com} + \hat{H}_{\text{int}} + \hat{H}_{\text{at-emf}} +\hat{H}_{\text{metric}}+ \hat{H}_{\text{mass defect}},
\end{equation}
where $\hat{H}_{\com}$ refers to COM and $\hat{H}_{\text{int}}$ to internal DOF only. The interaction of the atom with the external electromagnetic field is described by $\hat{H}_{\text{at-emf}}$. The relativistic coupling of internal and COM dynamics is covered by the two terms $\hat{H}_{\text{metric}}$ and $\hat{H}_{\text{mass defect}}$. In the following we will give the explicit form of all these terms in the relevant limit $m_1 \gg m_2$. 

The first contribution in~\eqref{eq:separatedH} is the COM Hamiltonian
\begin{align}\label{eq:Hcom}
	    \hat{H}_{\com}(M)=\hat{H}^{(0)}_{\com}+\hat{H}^{(1)}_{\com},
\end{align}
given by
\begin{align}
	    \hat{H}^{(0)}_{\com} &= \frac{\hat{\bar{\mathbf{P}}}^2}{2M}+ M \phi(\hat{\mathbf{R}}),\label{eq:Hcom0}\\
	    \hat{H}^{(1)}_{\com}&=-\frac{1}{2M c^2}\left(\frac{\hat{\bar{\mathbf{P}}}^2}{2M}\right)^2
	    +\frac{2 \gamma + 1}{2 M c^2} \hat{\bar{\mathbf{P}}} \cdot \phi(\hat{\mathbf{R}}) \hat{\bar{\mathbf{P}}}\nonumber\\
	    &\quad+ (2 \beta - 1) \frac{M (\phi(\hat{\mathbf{R}}))^2}{2 c^2},\label{eq:Hcom1}
\end{align}
which are the nonrelativistic COM Hamiltonian and the leading relativistic corrections, respectively. We define the total rest mass ${M=m_1+m_2}$ and $\hat{\bar{\mathbf{P}}}=\hat{\mathbf{P}}-Q\Ah(\hat{\mathbf{R}})$. The explicit definition of the relativistically corrected COM variables $\mathbf{R}$ and $\mathbf{P}$ are given in Eq.~\eqref{eq:COMcoordinates}. The COM Hamiltonian in Eq.~\eqref{eq:Hcom} corresponds simply to the approximately relativistic Hamiltonian for a charged point particle of mass $M$ in a gravitational field, minimally coupled to the electromagnetic field. For later use the COM Hamiltonian in Eq.~\eqref{eq:Hcom} is written in the form $\hat{H}_{\com}(M)$ with the total mass $M$ as an argument. 

The second contribution in Eq.~\eqref{eq:separatedH} is the Hamiltonian of the internal DOF
\begin{align}
		\hat{H}_{\text{int}}&=\hat{H}^{(0)}_\mathrm{int}+\hat{H}^{(1)}_\mathrm{int}\label{eq:H_int},
\end{align}
where
\begin{align}
		\hat{H}^{(0)}_\mathrm{int}&=\frac{\hat{\mathbf{p}}^2}{2\mu}
		+\frac{e_1e_2}{4\pi\varepsilon_0}\frac{1}{\hat{r}},\label{eq:Hint0}\\
		\hat{H}^{(1)}_\mathrm{int}&=
		-\frac{1}{2\mu c^2}\left(\frac{\hat{\mathbf{p}}^2}{2\mu}\right)^2\nonumber\\ &+\frac{e_1e_2}{4\pi\varepsilon_0}\frac{1}{2\mu Mc^2}\left(\hat{\mathbf{p}}\cdot\frac{1}{\hat{r}}\hat{\mathbf{p}} +\left(\hat{\mathbf{p}}\cdot\hat{\mathbf{r}} \right)\frac{1}{\hat{r}^3} \left(\hat{\mathbf{r}}\cdot\hat{\mathbf{p}} \right) \right)\label{eq:Hint1}
\end{align}
are the nonrelativistic Hamiltonian comprising the kinetic energy and  Coulomb energy and their lowest-order relativistic corrections, respectively. Here $\hat{\mathbf{p}}$ is the momentum associated with $\hat{\mathbf{r}}$ and $\mu$ is the reduced mass. Diagonalization of this Hamiltonian determines the electronic energy levels of the atom. Of course, in a complete description, other well-known relativistic corrections (e.g., concerning spin) will contribute to this Hamiltonian too. In the following, we assume these to be taken into account in the diagonalization of $\hat{H}_\mathrm{int}$.

The interaction of the atom with the electromagnetic field in the dipole approximation is given by
\begin{align}
		\hat{H}_{\text{at-emf}}&=Q\,\Phi(\hat{\mathbf{R}} )-\hat{\mathbf{d}}\cdot{\mathbf{E}}(\hat{\mathbf{R}})\nonumber\\
		&+\frac{1}{2M}\left[ \hat{\bar{\mathbf{P}}}\cdot\left( \hat{\mathbf{d}}\times{\mathbf{B}}(\hat{\mathbf{R}}) \right)+\mathrm{H.c.}\right]+\hat{H}_\mathrm{other},\label{HAFmaintext}
\end{align}
where $\hat{\mathbf{d}} = \sum_{i=1,2}e_i\left(\hat{\mathbf{r}}_i - \hat{\mathbf{R}} \right) $ is the electric dipole moment. The first line includes the electric potential and dipole energy and the second line represents the minimally coupled R\"ontgen term~\cite{Rontgen}. We suppress here further contributions involving the electromagnetic fields in $\hat{H}_\mathrm{other}$ which are given explicitly in Appendix~\ref{AppendixrelHam} and are not effective in the configuration of a Paul trap.

Finally, the last two terms of Eq.~\eqref{eq:separatedH}, $\hat{H}_{\text{metric}}$ and $\hat{H}_\text{mass defect}$, describe the relativistic coupling of COM and internal DOF and thus are the pivotal points of the following discussion. The first of these two terms is
\begin{align}\label{eq:Hmetric}
	    \hat{H}_\mathrm{metric}&= \gamma\frac{\phi(\hat{\mathbf{R}})}{c^2} \left(2  \frac{\hat{\mathbf{p}}^2}{2 \mu} +  \frac{e_1 e_2}{4 \pi \varepsilon_0} \frac{1}{\hat{r}}\right).
\end{align}
It is of metric origin and a consequence of space-time curvature, as is evident already from its proportionality to the PPN parameter $\gamma$ in the Eddington-Robertson metric in Eq.~\eqref{eq:metric}. We remind the reader that $\gamma=1$ in general relativity. More formally, the term follows when the Hamiltonian for the internal DOF is written in terms of distances measured with respect to the metric given in Eq.~\eqref{eq:metric} and expanded up to  $\mathcal{O}(c^{-4})$,
\begin{multline*}
    \frac{g_{ij}^{-1}(\hat{\mathbf{R}})\hat{p}_i\hat{p}_j}{2\mu}
		+\frac{e_1e_2}{4\pi\varepsilon_0}\frac{1}{\sqrt{g_{ij}(\hat{\mathbf{R}})\hat{r}_i\hat{r}_j}}\\
		\simeq\hat{H}_\mathrm{int}^{(0)}+\hat{H}_\mathrm{metric}.
\end{multline*}
Here summations run only over the spatial indices $i,j=1,2,3$. We refer to the recent work of Zych \textit{et al}.~\cite{Zych2019} for a more detailed discussion and references to previous literature discussing the metric correction $\hat{H}_\mathrm{metric}$. 

As a consequence of the virial theorem, the metric correction~\eqref{eq:Hmetric} turns out to be purely off-diagonal in the basis of stationary states with respect to the internal Hamiltonian $\hat{H}_{\text{int}}$ in Eq.~\eqref{eq:H_int}. This can be inferred from the identity
\begin{align}
\frac{\mathrm{i}}{\hbar}\left[\hat{\mathbf{r}}\cdot\hat{\mathbf{p}},H_{\text{int}}\right]=2  \frac{\hat{\mathbf{p}}^2}{2 \mu} +  \frac{e_1 e_2}{4 \pi \varepsilon_0} \frac{1}{\hat{r}}+\mathcal{O}(c^{-2}).
\end{align}
Since the average of the left-hand side with respect to eigenstates of $\hat{H}_{\text{int}}$ vanishes, the same holds for the right-hand side and thus also for Eq.~\eqref{eq:Hmetric}. Further constraints on the off-diagonal matrix elements can be concluded from noting that $\hat{H}_\mathrm{metric}$ is rotationally invariant. We do not go into further detail on this since the off-diagonal form of the metric term makes it ineffective as far as energy-nondegenerate states are concerned. For this case it can be neglected in a rotating-wave approximation with corrections scaling as $c^{-4}$. 

The second term describing relativistic coupling of COM and internal DOF is
\begin{align}\label{eq:Hmassdefect}
	    \hat{H}_{\text{mass defect}} &= \bigg(M \phi(\hat{\mathbf{R}})-\frac{\hat{\bar{\mathbf{P}}}^2}{2M} \bigg) \otimes\frac{\hat{H}_\mathrm{int}^{(0)}}{Mc^2}.
\end{align}
It can be interpreted as a result of the mass defect of the COM DOF due to the binding energy of the internal DOF. This interpretation is supported formally by the observation that, up to correction of $\mathcal{O}(c^{-4})$, the term $\hat{H}_{\text{mass defect}}$ can be absorbed in the COM Hamiltonian
\begin{align}\label{eq:HcomRel}
    \hat{H}_{\com}(M)+\hat{H}_{\text{mass defect}}\simeq
    \hat{H}_{\com}(M+\tfrac{\hat{H}_\mathrm{int}}{c^2}).
\end{align}
We remind the reader that $\hat{H}_{\com}(M)$ is defined as a \textit{function} of $M$ in Eq.~\eqref{eq:Hcom}. Thus the COM mass $M$ is effectively corrected by the mass defect and replaced by $M+\frac{\hat{H}_\mathrm{int}}{c^2}$. We note that the expression on the right-hand side of Eq.~\eqref{eq:HcomRel} is often used as a justification of the mass defect term on the left-hand side as, e.g., in the context of ion clocks~\cite{Jan2019mdtdequiv,Yudin2018, Haustein2019}. We would like to stress that it is the left-hand side of Eq.~\eqref{eq:HcomRel} that justifies the right-hand side in order $c^{-2}$, which is derived \textit{ab initio} from Eq.~\eqref{eq:Lagrangian1_grav} following~\cite{Schwartz2019}.

When $\Ah(\hat{\mathbf{R}})=0$ an alternative and no less justified interpretation of $\hat{H}_{\text{mass defect}}$ is that of a shift of the internal energies due to the gravitational redshift and due to the second-order Doppler effect, as is evident from
\begin{align}
    \hat{H}_\mathrm{int}+\hat{H}_{\text{mass defect}} &\simeq \hat{H}_\mathrm{int}\otimes\bigg(1-\frac{\hat{\mathbf{V}}^2}{2c^2}+\frac{\phi(\hat{\mathbf{R}})}{c^2}\bigg),
\end{align}
up to order $\mathcal{O}(c^{-4})$, where $\hat{V}$ corresponds to the velocity of the COM. These corrections are often added on a semiclassical basis as a result of time dilation when transforming from the COM rest frame to the laboratory frame~\cite{Ludlow2015_review}. We emphasize that the relativistic correction~\eqref{eq:Hmassdefect} thus accounts equally and fully for the time dilation due to the gravitational redshift and the second-order Doppler effect. 

In the following, it will be useful to rewrite this in the form
\begin{align}
    \hat{H}_\mathrm{int}+\hat{H}_{\text{mass defect}} &\simeq
    \hat{H}_\mathrm{int}\otimes\bigg(1+\frac{1}{c^2}\frac{\partial \hat{H}_{\com}(M)}{\partial M}\bigg)\nonumber\\
    &= \frac{\hbar\omega_0}{2}\hat{\sigma}_z\otimes\bigg(1+\frac{\delta\hat{\nu}}{\nu_0}\bigg)\label{eq:HintHmassdefect},
\end{align}
which is also valid in the presence of a magnetic vector potential. In the last step we performed a two-level approximation by restricting the description to two stationary bound states $\ket{g}$ and $\ket{e}$ with (negative binding) energies $h\nu_g$ and $h\nu_e$, respectively, and a transition frequency $\omega_0=2\pi\nu_0=2\pi(\nu_e-\nu_g)$. In these eigenstates and energies, we consider the relativistic corrections from Eq.~\eqref{eq:Hint1} already included. We defined the operator corresponding to the fractional frequency shift
\begin{align}\label{eq:fracshiftop}
    \frac{\delta\hat{\nu}}{\nu_0}=\frac{1}{c^2}\frac{\partial\hat{H}_{\com}(M)}{\partial M}
\end{align}
and implicitly absorbed a constant energy offset in the internal Hamiltonian. In the next section we will show that relativistic corrections due to coupling of internal and external DOFs in precision spectroscopy and frequency metrology can be discussed entirely on the basis of the fractional frequency shift operator in Eq.~\eqref{eq:fracshiftop}.

\section{Relativistic coupling of internal and external DOF in ion clocks}
\label{section3}

In this section we consider an optical clock based on a single ion in a Paul trap. That means we have to apply the relativistically corrected Hamiltonian in Eq.~\eqref{eq:separatedH} to the specific case of a charged composite particle subject to several external fields: first an external time-dependent electric potential realizing the confinement $\Phi(\hat{\mathbf{R}},t)$, second a weak gravitational field $\phi(\hat{\mathbf{R}})$, and third pulsed laser fields $\mathbf{E}(\hat{\mathbf{R}},t)$ driving the internal transition. In a Paul trap there is no vector potential, so we can replace $\hat{\bar{\mathbf{P}}}$ by $\hat{\mathbf{P}}$ in Eq.~\eqref{eq:separatedH}. The resulting Hamiltonian for a Paul trap including the relevant relativistic corrections is
\begin{align}
    \hat{H}&=\hat{H}_{\com}(M)+\frac{\hbar\omega_0}{2}\hat{\sigma}_z\bigg(1+\frac{\delta\hat{\nu}}{\nu_0}\bigg)
    -\hat{\mathbf{d}}\cdot\mathbf{E}(\hat{\mathbf{R}},t),\label{eq:HamRamsey}
\end{align}
where the fractional frequency shift is given in Eq.~\eqref{eq:fracshiftop}, and we collect all terms referring to the COM DOF in
\begin{align}\label{eq:HcomPaul}
    \hat{H}_{\com}(M,t)&=\frac{\hat{\mathbf{P}}^2}{2M}+M\phi(\hat{\mathbf{R}})+Q\,\Phi(\hat{\mathbf{R}},t).
\end{align}
Here we neglect or suppress the following relativistic corrections. (i) Terms in Eq.~\eqref{eq:Hcom1} affecting the COM DOF only are negligible for the small COM velocities of a cold ion and will affect the internal DOF in order $1/c^4$ only. (ii) In contrast, the corresponding terms of the internal DOF in Eq.~\eqref{eq:Hint1} are significant and contribute to its fine structure. We consider these terms to be absorbed in the internal states and energies. (iii) The metric term in Eq.~\eqref{eq:Hmetric} is dropped in a rotating-wave approximation, as explained earlier. (iv) The Röntgen term in the atom-field interaction~\eqref{HAFmaintext} scales as $\mathbf{P}/Mc$ and is negligible for a cold ion. Furthermore, it merely rescales the Rabi frequency of the pulses in a Ramsey interrogation (to be discussed in the next section) and will be compensated by their proper calibration.

\subsection{Ramsey spectroscopy}
\label{Subsection:RamseySpectroscopy}

For frequency spectroscopy, we treat a Ramsey interrogation scheme~\cite{Wineland_squeezed_1994} which amounts to a sequence of unitary evolutions $\ket*{\psi_\mathrm{out}}=\hat{U}_R(\omegaL T_R )\hat{U}(T_R)\hat{U}_R(0)\ket*{\psi_\mathrm{in}}$ where $\ket*{\psi_\mathrm{out\ (in)}}$ is the final (initial) state of internal and external DOF. For now, we consider pure states without loss of generality. Here $\hat{U}_R(\varphi)$ 
denotes the unitary during a Ramsey laser pulse, where $\varphi$ is the laser phase with respect to the atomic reference. Since such a pulse can be considered instantaneous compared to the duration $T_R$ of the Ramsey interrogation time, we neglect the mass defect during a Ramsey pulse. Therefore, $\hat{U}_R(\varphi)=\exp(-i\frac{\pi}{2}\qty(-\cos\varphi \hat{\sigma}_y+\sin\varphi \hat{\sigma}_x))$ and the laser phase in the second Ramsey pulse is $\varphi=\omegaL T_R$. During Ramsey interrogation, the ion evolves in the dark according to $\hat{U}(T_R)=\exp(\smash{-i\hat{H}T_R})$ with the Hamiltonian in Eq.~\eqref{eq:HamRamsey} where $\mathbf{E}(\hat{\mathbf{R}},t)=0$. In Appendix~\ref{appRamsey} we provide details on how this is evaluated.

After the second Ramsey pulse, the ion's internal state population $\hat{\sigma}_z$ is measured with average $\ev*{\sigma^\mathrm{out}_z}=\ev{\hat{\sigma}_z}{\psi_\mathrm{out}}$ and deviation $\Delta \sigma^\mathrm{out}_z$. For repeated measurements at a particular value of the detuning $\DeltaL=\omega_0-\omegaL$ the variance of the inferred frequency deviation of the clock laser from the atomic reference follows as
\begin{align}\label{eq:Deltaomega2}
    &\left(\Delta \omega\right)^2=\frac{\left(\Delta \sigma^\mathrm{out}_z\right)^2}{\qty|\frac{\partial \langle \sigma^\mathrm{out}_z\rangle}{\partial \omega}|^2}=\frac{\cos^2\left[\left(\DeltaL +\langle\delta \hat{\omega}\rangle\right)T_R\right] \Delta \sigma_z^2}{\langle\sigma_z\rangle^2 T_R^2 \sin^2\left[\left(\DeltaL +\langle\delta \hat{\omega}\rangle\right)T_R\right]} \nonumber\\
    &+\frac{\sin^2\left[\left(\DeltaL +\langle\delta \hat{\omega}\rangle\right)T_R\right] \Delta \sigma_y^2+T_R^2u^2\sin^2\left(  \DeltaL T_R\right)}{\langle\sigma_z\rangle^2 T_R^2 \sin^2\left[ \left(\DeltaL +\langle\delta \hat{\omega}\rangle\right)T_R\right]}.
\end{align}
This formula generalizes Eq.~(20) from Ref.~\cite{Wineland_squeezed_1994} for the mass defect. We define the mass defect (time dilation) shift in angular frequency $\delta \hat{\omega}=2\pi\delta\hat{\nu}$ and  the variance associated with the mass defect $u^2=\langle\delta\hat{\omega}^2\rangle- \langle\delta\hat{\omega}\rangle^2$. Averages in Eq.~\eqref{eq:Deltaomega2} concerning the internal DOF are taken with respect to the initial state $\ket{\psi_\mathrm{in}}$. In standard Ramsey spectroscopy, $\expval{\sigma_z}=-1$ and therefore $\Delta \sigma_z^2=0$ and $\Delta \sigma_y^2=1$. Furthermore, the expression for the inferred frequency deviation holds in leading (quadratic) order of the mass defect shift in both the numerator and denominator of Eq.~\eqref{eq:Deltaomega2}. Averages of $\hat{\omega}$ have to be understood with respect to the COM state, averaged over the interrogation time in the interaction picture with respect to the COM Hamiltonian, that is, Eq.~\eqref{eq:HcomPaul}.
This time average makes a nontrivial contribution only when the COM is in a nonstationary state and drops out when it is in a stationary, e.g., thermal, state.

Equation~\eqref{eq:Deltaomega2} implies that the Ramsey resonance
curve is shifted by $\expval{\delta\hat{\omega}}$ and exhibits a projection noise slightly increased by $u$. Thus, in order to evaluate the magnitude and relevance of these effects it is sufficient to consider the statistics of the operator corresponding to the fractional frequency shift in Eq.~\eqref{eq:fracshiftop} with respect to a given COM state. We note that the Feynman-Hellmann theorem can be conveniently used when evaluating the average systematic frequency shift with respect to an eigenstate $\ket{\psi}$ of the COM Hamiltonian with eigenenergy $E_\psi(M)$,
\begin{align}\label{eq:fracshift}
    \ev{\frac{\delta\hat{\nu}}{\nu_0}}_{\!\psi}=\frac{1}{c^2}\frac{\partial E_\psi\left(M\right)}{\partial M}.
\end{align}

\subsection{Quantum theory of an ion trap}\label{Subsection:Quantumtheoryofaniontrap}

For a rigorous discussion of the statistics of the fractional frequency shift, we briefly recapitulate the quantum theory of an ideal Paul trap following closely the notation of Leibfried \textit{et al}.~\cite{Leibfried2003_review} and adapting the quantum-mechanical treatment by Glauber~\cite{glauber1992}. Further below we will consider also corrections to the potentials in an ideal Paul trap (such as spurious dc electric fields and gravitational sack).

The potentials of a Paul trap contain direct current and alternating current components $\Phi(\hat{\bf{R}},t)= \Phi_{\text{dc}}(\hat{\bf{R}})+\Phi_{\text{ac}}(\hat{\bf{R}}, t),$ which at the trap center have the form of quadrupole fields 
\begin{align}\label{eq:trappotential}
		\Phi_{\text{dc}}(\hat{\bf{R}}) +\Phi_{\text{ac}}(\hat{\bf{R}}, t)= \frac{1}{2}\hat{\bf{R}}^T U\, \hat{\bf{R}} +\frac{1}{2} \cos(\Omega t) \hat{\bf{R}}^T \tilde{U}\, \hat{\bf{R}}.
\end{align}
Here $\Omega$ corresponds to the trap frequency and $U$ and $\tilde{U}$ are the dc and ac components of the quadrupole field tensors, respectively, and therefore correspond in general to symmetric traceless matrices. For an ideal trap geometry they are diagonal, $U=U_0\mathrm{diag}\qty(\alpha_1,\alpha_2,\alpha_3)$ and $\tilde{U}=\tilde{U}_0\mathrm{diag}\qty(\alpha'_1,\alpha'_2,\alpha'_3)$ with dimensionless coefficients $\alpha_i$ and $\alpha'_i$. The resulting COM Hamiltonian (neglecting gravity for the moment)
\begin{align}\label{eq:COMtrap}
    \hat{H}_\com(M,t)&=\frac{\hat{\mathbf{P}}^2}{2M}+Q\Phi(\hat{\mathbf{R}},t)
\end{align}
is explicitly time dependent and periodic with period $T=2\pi/\Omega$. We assume a stable trap configuration which supports quasistationary eigenenergy states $\ket{\mathbf{n},t}$ satisfying the generalized eigenvalue problem \cite{zel1967quasienergy,ritus1967shift,Sambe}
\begin{align}\label{eq:eigenvaleq}
    \qty(\hat{H}_{\com}(M,t)-\ii\hbar\partial_t)\ket{\mathbf{n},t}=E_\mathbf{n}(M)\ket{\mathbf{n},t}.
\end{align}
For the reader's convenience we summarize the solution of this eigenvalue problem in Appendix~\ref{apptrap}. The states $\ket{\mathbf{n},t}$ labeled by $\mathbf{n}=(n_1,n_2,n_3)$ are $T$-periodic Fock states whose time dependence accounts for the micromotion. The corresponding eigenenergies are
\begin{align}\label{eq:energylevels}
    E_\mathbf{n}(M)&=\sum_{i=1}^3\hbar\omega_i(M)\qty(n_i+\frac{1}{2}).
\end{align}
The motional eigenfrequencies, or trapping frequencies, are $\omega_i(M)=\frac{\Omega\beta_i(M)}{2}$, where $\beta^2_i(M)\simeq \mathtt{a}_i(M)+\frac{\mathtt{q}^2_i(M)}{2}$ is determined by the dimensionless Mathieu parameters $\mathtt{a}_i(M)=\frac{4QU_0\alpha_i}{M\Omega^2}$ and $\mathtt{q}_i(M)=\frac{2Q\tilde{U}_0\alpha'_i}{M\Omega^2}$ in lowest order $|\mathtt{a}_i|,\mathtt{q}_i^2\ll 1$. 

In the following, it will be important to note that the generalized Hamiltonian on the left-hand side of Eq.~\eqref{eq:eigenvaleq} has to be considered as acting on an enlarged Hilbert space \mbox{$\mathcal{H}_T=L_2(\mathds{R}^3)\otimes\mathcal{T}$}, where $\mathcal{T}$ is the space of $T$-periodic functions of time. We take care to construct the quasienergy eigenstates $\ket{\mathbf{n},t}$ within this space. As shown in Appendix~\ref{apptrap}, the Fock states are generated in the usual way by acting on a ground state $\ket{\mathbf{0},t}\in\mathcal{H}_T$ with creation operators which are adjoint to the annihilation operators
\begin{align}\label{eq:annihil}
    \hat{a}_i(t)=\frac{\ii\e^{-\ii\omega_i t}}{\sqrt{2\hbar M\omega_i}}\qty(u_i(t) \hat{P}_i-M \dot{u}_i(t) \hat{R}_i).
\end{align}
This expression is given in lowest order $|\mathtt{a}_i|,\mathtt{q}_i^2\ll 1$. Here both $\e^{-\ii\omega_i t}u_i(t)\simeq\qty(1+\frac{\mathtt{q}_i}{2}\cos(\Omega t))/\qty(1+\frac{\mathtt{q}_i}{2})$ and $\e^{-\ii\omega_i t}\dot{u}_i(t)$ are $T$-periodic functions. These operators obey equal-time bosonic commutation relations $[\hat{a}_i(t),\hat{a}^\dagger_j(t)]=\delta_{ij}$. The ground state is determined by  $\hat{a}_i(t)\ket{\mathbf{0},t}=0$ for $i=1,2,3$. We refer the reader to Appendix~\ref{apptrap} for details.

When evaluating the effects of relativistic corrections it will be necessary to calculate matrix elements of $T$-periodic operators with respect to $T$-periodic states. The corresponding scalar products have to be understood within the enlarged Hilbert space $\mathcal{H}_T$, and therefore involve a time average over one period $T$. In order to indicate this explicitly, we write this scalar product as $\llangle\psi|\varphi\rrangle=\frac{1}{T}\int_0^T\dd{t}\braket{\psi(t)}{\varphi(t)} $ for $\ket{\psi},\ket{\varphi}\in\mathcal{H}_T$, where $\braket{\psi(t)}{\varphi(t)}$ is the usual scalar product in $L_2(\mathds{R}^3)$. Accordingly, the average of an operator $A(t)$ with respect to a state $\ket{\psi}\in\mathcal{H}_T$ has to be understood as
\begin{align}
    \llangle{A}\rrangle_\psi=\frac{1}{T}\int_0^T\dd{t}\ev{A(t)}{\psi(t)},   
\end{align}
where $\ev{A(t)}{\psi(t)}$ is the average value in $L_2(\mathds{R}^3)$.

\subsection{Fractional frequency shift due to mass defect: second order Doppler effect}

We are now ready to evaluate the fractional frequency shift \eqref{eq:fracshiftop} due to the mass defect for a given COM state. Here we will consider in particular Fock states $\ket{\mathbf{n},t}$ and thermal mixtures of Fock states at (pseudo)temperatures $T_i$ for motion along axis $i$. However, the treatment is general and can be applied just as well to any other quantum state. In view of Eqs.~\eqref{eq:COMtrap} and \eqref{eq:fracshiftop}, the fractional frequency shift of a trapped ion is entirely due to its kinetic COM energy since
\begin{align}
    \frac{\delta\hat{\nu}}{\nu_0}&=-\frac{\hat{K}}{Mc^2}, &
    \hat{K}&=\frac{\hat{\mathbf{P}}^2}{2M}\label{eq:fracshiftkinetic}.
\end{align}
The operator for the kinetic energy can be expressed in terms of creation and annihilation operators by inversion of Eq.~\eqref{eq:annihil}. In this way, mean values and uncertainties can be easily evaluated, and the only integrals that remain to be calculated are those over the period $T$ of micromotion.

For the average fractional shift due to a COM Fock state $\ket{\mathbf{n},t}$ the Feynman-Hellmann theorem can be applied. With Eqs.~\eqref{eq:fracshift} and \eqref{eq:energylevels} we find
\begin{align}\label{eq:fracshiftFock}
    \llangle[\Big]{\frac{\delta\hat{\nu}}{\nu_0}}\rrangle[\Big]_{\!\mathbf{n}}=
    -\sum_{i=1}^3\frac{\hbar\omega_i\qty(n_i+\frac{1}{2})}{2Mc^2}\left(1+\frac{\mathtt{q}_i^2}{2\mathtt{a}_i+\mathtt{q}_i^2}\right),
\end{align}
in leading order of $\mathtt{a}_i$ and $\mathtt{q}_i^2$. We refer the reader to Appendix~\ref{apptrap} for remarks on how this expansion is technically performed here and in all subsequent formulas. Here the first term corresponds to the secular motion and the last one to the micromotion. The fractional shift for a thermal state with average occupation numbers $\bar{\mathbf{n}}=(\bar{n}_1,\bar{n}_2,\bar{n}_3)$ will have the same form as Eq.~\eqref{eq:fracshiftFock} where the Fock state number $n_i$ is replaced by $\bar{n}_i=1/\big(\exp\big(\frac{\hbar\omega_i}{k_BT_i}\big)-1\big)$. In the high-temperature limit $\bar{n}_i\approx \frac{k_BT_i}{\hbar \omega_i}$ we arrive at  
\begin{align}
\llangle[\Big]{\frac{\delta\hat{\nu}}{\nu_0}}\rrangle[\Big]_{\!\bar{\mathbf{n}}}=
    -\sum_{i=1}^3\frac{k_BT_i}{Mc^2}\frac{\mathtt{a}_i+\mathtt{q}_i^2}{2\mathtt{a}_i+\mathtt{q}_i^2}.
\end{align}
This recovers the result of Berkeland \textit{et al}.~\cite{Berkeland1998} [cf. the first term on the right-hand side of Eq. (30)] on the second-order Doppler (time-dilation) shift for a thermal state. In the opposite limit $k_BT_i\ll\hbar\omega_i$, zero-point fluctuations in both secular and micromotion in the quantum ground state still cause a fractional shift, as follows from Eq.~\eqref{eq:fracshiftFock} for $n_i=0$. Calculating the average of Eq.~\eqref{eq:fracshiftkinetic} directly, that is $\llangle\delta\hat{\nu}/\nu_0\rrangle=-\llangle\hat{K}\rrangle/Mc^2$,  using the algebra of creation and annihilation operators defined in Eq.~\eqref{eq:annihil} yields the same result in leading order of $\mathtt{a}_i$ and $\mathtt{q}_i^2$. 

So far, we considered an ideal quadrupole potential for the trap in the form of Eq.~\eqref{eq:trappotential}.  In reality various deviations from this ideal geometry will occur and impact the second-order Doppler shift. In the following sections we will consider additional linear potentials due to uncompensated dc electric fields (Sec.~\ref{sec:dcforces}) and gravity (Sec.~\ref{sec:gravity}), as well as spurious electric quadrupole fields (Sec.~\ref{sec:quadrupole}).

\subsection{Fractional frequency shift due to dc forces and excess micromotion}\label{sec:dcforces}

Here we consider an additional linear dc electric potential due to uncompensated stray fields~\cite{Berkeland1998} causing so-called excess micromotion. We include such a spurious potential by adding to the COM Hamiltonian a perturbation
\begin{align}\label{eq:DCpert}
    \hat{H}_\mathrm{dc}=-Q\mathbf{E}_\mathrm{dc}\cdot\hat{\mathbf{R}},
\end{align}
where $\mathbf{E}_\mathrm{dc}=(\mathrm{E}_{\mathrm{dc},1},\mathrm{E}_{\mathrm{dc},2},\mathrm{E}_{\mathrm{dc},3})$ is the dc electric field at the trap center. 

This problem can still be solved exactly. Just as in the previous sections, we assume a stable trap configuration which supports quasi-stationary eigenenergy states $\ket{\mathbf{n},t}_\mathrm{dc}$ satisfying the generalized eigenvalue problem 
\begin{align}
    \qty(\hat{H}_{\com}(M,t)-\ii\hbar\partial_t+\hat{H}_\mathrm{dc})\ket{\mathbf{n},t}_\mathrm{dc}=E^\mathrm{dc}_\mathbf{n}(M)\ket{\mathbf{n},t}_\mathrm{dc}.
\end{align}
The modified creation operators generating the Fock states $\ket{\mathbf{n},t}_\mathrm{dc}$ are derived in Appendix~\ref{apptrap}. The modified eigenenergies are $E^\mathrm{dc}_\mathbf{n}(M)=E_\mathbf{n}(M)+E_\mathrm{dc}(M)$, where the energy correction due to the dc field  does not depend on the Fock number $\mathbf{n}$ and is given by
\begin{align}
    E_{\mathrm{dc}}(M)
    &=-\sum_{i=1}^3\frac{4\mathrm{E}^2_{\mathrm{dc},i}Q^2}{M\qty(2\mathtt{a}_i(M)+{\mathtt{q}_i^2(M)})\Omega^2}.
\end{align}
We note that this result can also be found by accounting for the perturbation in Eq.~\eqref{eq:DCpert} in second-order perturbation theory. This is due to the fact that the perturbation \eqref{eq:DCpert} is time independent and linear in position and the unperturbed Hamiltonian is quadratic in position and momentum. Therefore, the exact energy eigenstates are suitably displaced Fock states and the quasi energies will be shifted by a constant quadratic in the perturbation.

Thus, the dc electric field causes a fractional frequency shift due to the excess micromotion that can be evaluated as before using the Feynman-Hellmann theorem. On top of the thermal shift in Eq.~\eqref{eq:fracshiftFock}, the dc field adds a shift
\begin{align}
    \llangle[\Big]{\frac{\delta\hat{\nu}}{\nu_0}}\rrangle[\Big]_{\!\mathrm{dc}}&=\frac{1}{c^2}\pdv{E_\mathrm{dc}(M)}{M}\nonumber\\
    &=-\sum_{i=1}^3\qty(\frac{2\mathtt{q}_i\mathrm{E}_{\mathrm{dc},i}Q}{Mc\qty(2\mathtt{a}_i+{\mathtt{q}_i^2})\Omega})^2,\label{eq:fracshiftdc}
\end{align}
and reproduces again the result of Berkeland \textit{et al}.~\cite{Berkeland1998}. Here we have used the notation $\llangle \rrangle_{\mathrm{dc}}$, which emphasizes that the contribution is the same for all Fock states, independent of the index $\mathbf{n}$. The fractional frequency shift of a trapped ion in this case is still entirely due to its kinetic COM energy as shown in Eq.~\eqref{eq:fracshiftkinetic}. With the modified creation and annihilation operators one can easily verify Eq.~\eqref{eq:fracshiftdc} via  $\llangle\delta\hat{\nu}/\nu_0\rrangle=-\llangle\hat{K}\rrangle/Mc^2$ in leading order of $\mathtt{a}_i$ and $\mathtt{q}_i^2$.

\subsection{Fractional frequency shift due to gravity}\label{sec:gravity}

In this section we consider a contribution to the fractional frequency shift due to an interplay between position fluctuations and gravitational redshift, as discussed in~\cite{Haustein2019}. The coupling to the gravitational field is described by adding to the COM Hamiltonian
\begin{align}
    \hat{H}_\mathrm{g}=M\phi(\hat{\mathbf{R}})\label{eq:gravitytrap},
\end{align}
where we approximate the gravitational potential in linear order $\phi(\hat{\mathbf{R}})=\phi_0+\mathbf{g}\cdot\hat{\mathbf{R}}$. Here, $\phi_0$ is the gravitational potential at the trap center and $\mathbf{g}=(g_1,g_2,g_3)$. In this  linear approximation, the effect of gravity can again be taken into account with the accordingly modified creation operators [cf. Eq. \eqref{eq:alpha}]. 

Including gravity, the fractional frequency shift of a trapped ion is due to both its kinetic COM energy and the contribution due to gravity. On top of the thermal shift in Eq.~\eqref{eq:fracshiftFock}, gravity adds a shift
\begin{align}\label{eq:deltanugrav}
    \frac{\delta\hat{\nu}}{\nu_0}&=-\frac{\hat{K}}{Mc^2}+\frac{\phi_0+\mathbf{g}\cdot\hat{\mathbf{R}}}{c^2}.
\end{align}
This follows from Eqs.~\eqref{eq:fracshiftop}, \eqref{eq:COMtrap}, and \eqref{eq:gravitytrap}. In this case we can write the fractional frequency shift as $\llangle{\frac{\delta\hat{\nu}}{\nu_0}}\rrangle=\llangle{\frac{\delta\hat{\nu}}{\nu_0}}\rrangle_{\!\mathbf{n}}+\llangle{\frac{\delta\hat{\nu}}{\nu_0}}\rrangle_{\!\mathrm{g}}$, where the first contribution is given in Eq.~\eqref{eq:fracshiftFock} and
\begin{align}\label{eq:fracshiftgravity}
    \llangle[\Big]{\frac{\delta\hat{\nu}}{\nu_0}}\rrangle[\Big]_{\!\mathrm{g}}=-\sum_{i=1}^3
    \frac{4\mathtt{a}_i+3\mathtt{q}_i^2}{4\mathtt{a}_i+2\mathtt{q}_i^2}
    \frac{g^2_i}{\omega^2_ic^2}+\frac{\phi_0}{c^2}.
\end{align}
Here we have used the notation $\llangle \rrangle_{\mathrm{g}}$ to denote the shift due to gravity. Due to the linearity in $\hat{\mathbf{R}}$ of the second term in Eq.~\eqref{eq:deltanugrav}, this shift is affecting all Fock states in the same way and therefore is independent of temperature for thermal states. In the case of a dc harmonic confinement $\mathtt{q}_i\rightarrow 0$ and neglecting the background redshift $\phi_0=0$, we recover the result of Haustein \textit{et al}.~\cite{Haustein2019}. The results from the two preceding sections can be easily combined to take into account both the dc forces and gravity, which will lead to some cross terms between the two effects as shown in Appendix~\ref{apptrap}.

\begin{figure}[t]
\centering
\includegraphics{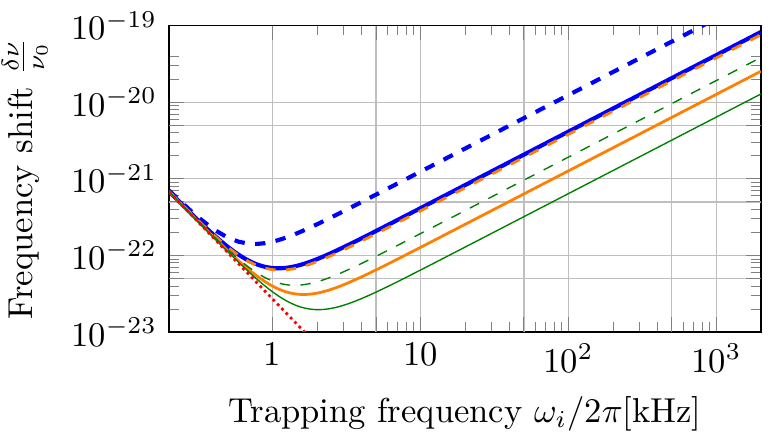}
\caption{Redshift $\frac{g^2}{\omega_i^2 c^2}$ (red dotted line) and total fractional frequency shift, i.e., redshift plus second-order Doppler shift $\frac{\hbar\omega_i(\bar{n}+1/2)}{2Mc^2}$, for $\bar{n}=0$ (solid lines) and $\bar{n}=1$ (dashed lines) for Al$^+$ (thick blue lines), Yb$^+$ (thin green lines), and neutral Sr (thick orange lines) versus trapping frequency $\omega_i$. We consider here a static confinement $\mathtt{q}_i=0$ for simplicity.}
\label{fig:frequencyshift}
\end{figure}

It is noteworthy that the fractional frequency shift due to the kinetic energy~\eqref{eq:fracshiftFock} grows linearly with the trap frequency (corresponding to an increased kinetic energy for tighter trapping), while the trap-dependent contribution due to gravity~\eqref{eq:fracshiftgravity} decreases quadratically with the trap frequency due to a decreased fluctuation in position and therefore also in potential energy. The trade-off with respect to the trap frequency has been discussed and optimized by Haustein \textit{et al}. \cite{Haustein2019} in order to minimize the fractional frequency shift. For the parameter regime of a conventional ion trap, the redshift term \eqref{eq:fracshiftgravity} will be smaller than the second-order Doppler shift \eqref{eq:fracshiftFock} (see Fig.~\eqref{fig:frequencyshift}). With the present formulas, it is straightforward to extend this discussion to account for micromotion. The results are unwieldy and will be reported elsewhere~\cite{Martinez_thesis}. 

\subsection{Variance of the fractional frequency shift}

In the preceding section we considered the \textit{average} fractional frequency shift which enters the inferred frequency deviation in Eq.~\eqref{eq:Deltaomega2} as a systematic shift of the Ramsey resonance curves. Now we will address the role of the \textit{uncertainty} in the frequency shift $u^2=\langle\delta\hat{\omega}^2\rangle- \langle\delta\hat{\omega}\rangle^2$, which has been suggested~\cite{Sinha2014} to pose a fundamental limitation to the precision of an optical clock. For the case of standard Ramsey interrogation (where $\expval{\sigma_z}=-1$ and therefore $\Delta \sigma_z^2=0$ and $\Delta \sigma_y^2=1$) and taking into account $N$ independent interrogations,  Eq.~\eqref{eq:Deltaomega2} can be rewritten for the inferred relative frequency deviation \begin{align}\label{eq:Deltaomega2simple}
    \qty(\frac{\Delta \omega}{\omega_0})^2&=\frac{1}{N}\qty(\frac{1}{\omega^2_0T_R^2}+\frac{u^2}{\omega^2_0}),
\end{align}
which holds to leading order in $u^2$. The first term on the right-hand side is the projection noise and the second term accounts for the variance of the fractional frequency shift due to the mass defect. 

By means of the ladder operators \eqref{eq:annihil}, it is straightforward to evaluate the latter, and we find for a thermal COM state 
\begin{align}    \label{eq:deviation}
    \frac{u^2}{\omega_0^2}&=\frac{\llangle\hat{K}^2\rrangle_{\bar{\mathbf{n}}}-\llangle\hat{K}\rrangle_{\bar{\mathbf{n}}}^2}{M^2_0c^4}\nonumber\\
    &=\sum_{i=1}^3
    2\qty(\frac{\hbar\omega_i(\bar{n}_i+\frac{1}{2})}{2Mc^2})^2\\
    &\quad\times\left[\qty(1+\frac{\mathtt{q}_i^2}{2\mathtt{a}_i+\mathtt{q}_i^2})^2+\frac{3}{4}\frac{\mathtt{q}_i^4}{\qty(2\mathtt{a}_i+\mathtt{q}_i^2)^2}\right].\nonumber
\end{align}
This expression is given in leading order of Mathieu parameters, taking into account one order more than in other cases as we are showing a variance and not the standard deviation. The standard deviation associated with this variance can be interpreted as the quantum fluctuations of the kinetic energy $\Delta \hat{K}/Mc^2$. For evaluating the second moment of the kinetic energy it is convenient to use $\langle\hat{\mathbf{P}}^4\rangle_{\bar{\mathbf{n}}}=3\langle\hat{\mathbf{P}}^2\rangle^2_{\bar{\mathbf{n}}}$ for Gaussian statistics. Note that this identity only holds for the average in $L_2(\mathds{R}^3)$, as the statistics with respect to time are non-Gaussian. We caution that the fractional frequency uncertainty for a COM Fock state (which is non-Gaussian) looks slightly different but can still be easily evaluated by means of creation and annihilation operators. 

It is interesting to consider in Eq.~\eqref{eq:deviation} the contributions along different directions $i$ in the limit of a pure dc or ac potential. For a dc potential $\mathtt{q}_i=0$, one finds for a thermal COM state that the standard deviation in the direction $i$ corresponds to $\sqrt{2}$ times the fractional frequency shift in the same direction. This agrees with the expectation that without micromotion the Gaussian statistics entails a variance $\text{Var}(\hat{K})_i=2\llangle\hat{K}_i\rrangle^2$ for the kinetic energy along this direction. For a pure ac potential $\mathtt{a}_i=0$, one finds instead a standard deviation in the direction $i$ of the fractional frequency shift of $\sqrt{19/8}$ times the fractional frequency shift in the same direction. The slight increase in comparison to a dc potential is due to micromotion. In the same way, the uncertainty in the fractional frequency shift can be evaluated for the contribution of the excess micromotion \eqref{eq:fracshiftdc}.

The standard deviation of the fractional frequency shift $u/\omega_0$ implied by Eq.~\eqref{eq:deviation} has the same order of magnitude as the \textit{average} second-order Doppler fractional frequency shift. However, in order to resolve the latter, a large number of measurements $N$ is required to average down the projection noise [first term in Eq.~\eqref{eq:Deltaomega2simple}] to the level of the systematic second-order Doppler shift. It is important to note that in the same course the uncertainty of the second-order Doppler shift is suppressed by $N$. Thus, its contribution to Eq.~\eqref{eq:Deltaomega2simple} should not be misinterpreted to imply a fundamental limit to the stability of an ion clock. Rather, the standard deviation of the fractional frequency shift should be considered as a (relatively small) correction to the quantum projection noise of a single measurement on a two-level system. However, it does imply a limit to the short-term stability.

\subsection{Effect of additional quadrupole fields}\label{sec:quadrupole}

Finally, we address the effect of a spurious electric quadrupole field. Yudin and Taichenachev~\cite{Yudin2018} suggested that an additional quadrupole field beyond the ideal dc and ac potentials in Eq.~\eqref{eq:trappotential}, in interplay with the mass defect, could lead to systematic shifts that have not been considered before. Compared to Eq.~\eqref{eq:trappotential}, an additional quadrupole electric field can manifest itself in a shift of the minima between the ac and dc potentials, a change in their curvature, and/or a shift in their axes. 

In order to take this into account, we choose, without loss of generality, the origin of our coordinates to coincide with the zero point of the ac contribution and the coordinate basis to be aligned with its axes, i.e., with the eigenvectors of $\tilde{U}$ in Eq.~\eqref{eq:trappotential}. This leaves the ac contribution unchanged from the previous sections. The \emph{total} (intended and accidental) dc potential is now expressed in an expansion around the origin,
\begin{equation}
\Phi_{\text{dc}} (\hat{\bf{R}}) =  \Phi_{\text{dc}}\left(\hat{\bf{0}}\right) + \mathbf{E}_\mathrm{dc}\cdot\hat{\bf{R}} + \frac{1}{2}\hat{\bf{R}}^T U\, \hat{\bf{R}},
\end{equation}
where $U_{ij}=\frac{\partial^2 \Phi_{\text{dc}}(\hat{\bf{0}})}{\partial R_i \partial R_j}$ and $\hat{\bf{0}}$ stands for the center of the ac potential. The $\Phi_{\text{dc}}\left(\hat{\bf{0}}\right)$ will be a constant shift in the Hamiltonian and therefore will have no effect. The linear component can be interpreted as a contribution to the spurious $\mathbf{E}_\mathrm{dc}$ field studied before in Sec.~\ref{sec:dcforces}. Finally, without loss of generality we write $U=U_0\mathrm{diag}\qty(\alpha_1,\alpha_2,\alpha_3)+W$, with dimensionless coefficients $\alpha_i$ and a purely off-diagonal perturbation $W$. The diagonal terms determine the potential curvatures and hence the effective trap frequencies, as in the previous sections. A calibration of the second-order Doppler effect, based on trap spectroscopy and thermometry of the COM motion, will thus properly account for potential deviations of the $\alpha_i$ from their nominal values. It is these changes that were discussed in~\cite{Yudin2018} in a perturbative account. We thus agree with Yudin and Taichenachev that a change in potential curvature can enter the systematics in relevant magnitude, but notice that these effects are already taken into account in the operational calibration of an ion clock in the context of the second-order Doppler effect.

It remains to discuss the effect of axes misalignments. For this, we treat the nondiagonal correction as a perturbation to the Hamiltonian~\eqref{eq:COMtrap},
\begin{align}
    \hat{H}_\mathrm{off-diag}=Q\hat{\mathbf{R}}^T W\hat{\mathbf{R}}.
\end{align}
Assuming for simplicity a nondegenerate spectrum of motional eigenfrequencies $\omega_i$, we can employ nondegenerate perturbation theory \cite{Sambe} in order to evaluate the corrections to the energy levels of quasistationary states $\ket{\mathbf{n},t}$ in Eq.~\eqref{eq:energylevels}. Expressing the position operator $\hat{\mathbf{R}}$ in terms of creation and annihilation operators, it follows immediately (due to the off-diagonal nature of $W$) that the first-order correction vanishes $\llangle \hat{H}_\mathrm{off-diag}\rrangle_\mathbf{n}=0$. Of course, the eigenstates will change. In the case of nominal degeneracies in the trap frequencies the off diagonal terms can be treated as a perturbation on the level of the Mathieu equation, following the treatment shown by Landa \textit{et al.}~\cite{Landa2012}. This will lead to a lifting of the degeneracies, which will again be accounted for in the trap calibration.

\section{Conclusions}\label{sec:Conclusions}
In this article we have presented a systematic and fully quantum-mechanical treatment of relativistic frequency shifts in atomic clocks based on trapped ions. We started by deriving an approximate relativistic Hamiltonian for the center of mass and internal dynamics of an electromagnetically bound, charged two-particle system in external electromagnetic and gravitational fields. We applied this Hamiltonian to an ion in a Paul trap, including the effects of micromotion, excess micromotion, and trap imperfections. We recovered results known from semiclassical treatments based on time-dilation arguments. The Hamiltonian \textit{ab initio} treatment given here avoids the need for \textit{ad hoc} arguments based on time-dilation or mass-defect corrections. It provides a solid basis for applications to more complicated systems, such as atom clocks based on ion crystals as well as neutral lattice clocks. It would be desirable to account for spin, along the lines of~\cite{Pachucki2004,Pachucki2007} but taking into account the gravitational field.

\begin{acknowledgments}
We acknowledge support from Deutsche Forschungsgemeinschaft (DFG, German Research Foundation) under Germany’s Excellence Strategy – EXC-2123 QuantumFrontiers – 390837967, and Project-ID
274200144 – SFB 1227 (DQ-mat, project A05, A06 and B03). 
This project has received funding from the EMPIR programme cofinanced by the Participating States and from the European Unions Horizon 2020 research and innovation program. This work was supported by the EMPIR project 18SIB05 “Robust Optical Clocks for International Timescales”.
This work was also supported by the Max Planck-RIKEN-PTB-Center for Time, Constants and Fundamental Symmetries.
We thank Jan Kiethe and Marius Schulte for comments on early versions of the paper and Philip Schwartz and Domenico Giulini for discussions.
\end{acknowledgments}

\appendix

\section{Hamiltonian for ion in external electromagnetic fields with first order relativistic corrections}

\label{AppendixrelHam}

The derivation in Sec.~\eqref{section2} follows the one by Sonnleitner and Barnett \cite{Sonnleitner2018} and Schwartz and Giulini \cite{Schwartz2019}. In this appendix we highlight the main differences in the derivation with respect to these works and show the terms that are obtained for a charged composite particle system instead of a neutral one. Apart from the differences highlighted here, the steps of the calculations are the same and are not affected by having $Q\neq 0$. For details of the calculations, we refer the reader to \cite{Sonnleitner2018,Schwartz2019} as well as to \cite{Close1970,Lembessis1993}. The textbooks \cite{johndavidjackson1975,Ldlandau1980,CohenTannoudji1997} are useful sources for these types of calculations.

One of the main differences arises during the separation of COM and relative DOF involved in passing from Eq.~\eqref{eq:TwoBodyHamiltonian} to Eq. \eqref{eq:separatedH}. As explained in the main text, the definitions of internal and external coordinates, COM coordinate $\hat{{\mathbf{R}}}=(m_1\mathbf{r}_1+m_2\mathbf{r}_2)/M$, and relative coordinate $\hat{\mathbf{r}}=\hat{\mathbf{r}}_2-\hat{\mathbf{r}}_1$ with respective momenta $\hat{{\mathbf{P}}}$ and $\hat{\mathbf{p}}$, need to be adjusted in order to account for relativistic corrections. Suitable canonical transformations have been introduced by Close and Osborn \cite{Close1970} and employed in \cite{Sonnleitner2018} for neutral systems. For a composite system with net charge $Q\neq 0$, we generalize this transformation by seeking coordinates $\hat{\widetilde{\mathbf{R}}}$ and $\hat{\widetilde{\mathbf{r}}}$ with respective momenta $\hat{\widetilde{\mathbf{P}}}$ and $\hat{\widetilde{\mathbf{p}}}$ which fulfill
\begin{align}\label{eq:COMcoordinates}
&\hat{\mathbf{R}}=\hat{\widetilde{\mathbf{R}}}+\frac{m_1-m_2}{2 M^2 c^2}\left[\left(\frac{\hat{\widetilde{\mathbf{p}}}^2}{2\mu}\hat{\widetilde{\mathbf{r}}}+\text{H.c.} \right) +\frac{e_1e_2}{4\pi\varepsilon_0\hat{\widetilde{r}}}\hat{\widetilde{\mathbf{r}}} \right]\nonumber\\
&\quad \quad-\frac{1}{4M^2c^2}\left[\left(\hat{\widetilde{\mathbf{r}}}\cdot \hat{\widetilde{\mathbf{P}}}' \right)\hat{\widetilde{\mathbf{p}}}+\left(\hat{\widetilde{\mathbf{P}}}'\cdot \hat{\widetilde{\mathbf{p}}} \right)\hat{\widetilde{\mathbf{r}}} +\text{H.c.} \right], \nonumber \\
&\hat{\mathbf{P}}=\hat{\widetilde{\mathbf{P}}}+f\left(\hat{\widetilde{\mathbf{R}}},\hat{\widetilde{\mathbf{P}}},\hat{\widetilde{\mathbf{r}}},\hat{\widetilde{\mathbf{p}}}\right)\nonumber\\
&\hat{\mathbf{r}}=\hat{\widetilde{\mathbf{r}}}+\frac{m_1-m_2}{2\mu M^2 c^2}\left[\left(\hat{\widetilde{\mathbf{r}}}\cdot \hat{\widetilde{\mathbf{P}}}' \right)\hat{\widetilde{\mathbf{p}}} +\text{H.c.} \right]-\frac{\hat{\widetilde{\mathbf{r}}}\cdot\hat{\widetilde{\mathbf{P}}}'}{2M^2c^2}\hat{\widetilde{\mathbf{P}}}',\nonumber\\
&\hat{\mathbf{p}}=\hat{\widetilde{\mathbf{p}}}+\frac{\hat{\widetilde{\mathbf{p}}}\cdot\hat{\widetilde{\mathbf{P}}}'}{2M^2c^2}\hat{\widetilde{\mathbf{P}}}'-\frac{m_1-m_2}{2M^2c^2}\Biggl[\frac{\hat{\widetilde{\mathbf{p}}}^2}{\mu}\hat{\widetilde{\mathbf{P}}}'\nonumber\\
&\quad \quad+\frac{e_1e_2}{4\pi\varepsilon_0}\Biggl(\frac{1}{\hat{\widetilde{r}}}\hat{\widetilde{\mathbf{P}}}'-\frac{\hat{\widetilde{\mathbf{P}}}'\cdot \hat{\widetilde{\mathbf{r}}}}{\hat{\widetilde{r}}^3}\hat{\widetilde{\mathbf{r}}} \Biggr)  \Biggr].
\end{align}
Here we include a minimal coupling of the COM DOF to the electromagnetic field $\hat{\widetilde{\mathbf{P}}}'=\hat{\widetilde{\mathbf{P}}}-Q\Ah(\hat{\widetilde{\mathbf{R}}})$. In order to still have a canonical transformation, the definition of the COM momentum involves an ansatz function $f\left(\hat{\widetilde{\mathbf{R}}},\hat{\widetilde{\mathbf{P}}},\hat{\widetilde{\mathbf{r}}},\hat{\widetilde{\mathbf{p}}}\right)$. Enforcing canonical commutation relations $\left[\hat{\widetilde{\mathbf{R}}}_k,\hat{\widetilde{\mathbf{P}}}_l\right]=\left[\hat{\widetilde{\mathbf{r}}}_k,\hat{\widetilde{\mathbf{p}}}_l \right] =\ii\hbar \delta_{kl}$ and $\left[\hat{\widetilde{\mathbf{R}}}_k,\hat{\widetilde{\mathbf{r}}}_l\right]=\left[\hat{\widetilde{\mathbf{R}}}_k,\hat{\widetilde{\mathbf{p}}}_l\right]=\left[\hat{\widetilde{\mathbf{P}}}_k,\hat{\widetilde{\mathbf{p}}}_l\right]=\left[\hat{\widetilde{\mathbf{P}}}_k,\hat{\widetilde{\mathbf{r}}}_l\right]=0$, we find
\begin{align}\label{eq:equationf}
    &f\left(\hat{\widetilde{\mathbf{R}}},\hat{\widetilde{\mathbf{P}}},\hat{\widetilde{\mathbf{r}}},\hat{\widetilde{\mathbf{p}}}\right)\nonumber\\
    = &~\frac{m_1 - m_2}{2M^2c^2} Q \Biggl[\Biggl(\frac{\hat{\widetilde{\mathbf{p}}}^2}{2\mu} \nabla_{\hat{\widetilde{\mathbf{R}}}} \left(\hat{\widetilde{\mathbf{r}}} \cdot \Ah(\hat{\widetilde{\mathbf{R}}})\right) + \text{\text{H.c.}}\Biggr)\nonumber\\
    &+\frac{e_1 e_2}{4 \pi \varepsilon_0 \hat{\widetilde{r}}} \nabla_{\hat{\widetilde{\mathbf{R}}}} \left(\hat{\widetilde{\mathbf{r}}} \cdot \Ah(\hat{\widetilde{\mathbf{R}}})\right)\Biggr]\nonumber\\
	&-\frac{Q}{4M^2c^2} \left[\hat{\widetilde{\mathbf{r}}} \cdot \hat{\widetilde{\mathbf{P}}}' \nabla_{\hat{\widetilde{\mathbf{R}}}} \left(\hat{\widetilde{\mathbf{p}}} \cdot \Ah(\hat{\widetilde{\mathbf{R}}})\right)\nonumber\right.\\
	&\left.+ \hat{\widetilde{\mathbf{P}}}' \cdot \hat{\widetilde{\mathbf{p}}} \nabla_{\hat{\widetilde{\mathbf{R}}}} \left(\hat{\widetilde{\mathbf{r}}} \cdot \Ah(\hat{\widetilde{\mathbf{R}}})\right) + \text{\text{H.c.}}\right].
\end{align}
In principle, every function $g\left(\hat{\widetilde{\mathbf{R}}},\hat{\widetilde{\mathbf{P}}},\hat{\widetilde{\mathbf{r}}},\hat{\widetilde{\mathbf{p}}}\right) = f\left(\hat{\widetilde{\mathbf{R}}},\hat{\widetilde{\mathbf{P}}},\hat{\widetilde{\mathbf{r}}},\hat{\widetilde{\mathbf{p}}}\right) +h\left(\hat{\widetilde{\mathbf{R}}}\right)$ that fulfills $\frac{\partial h_j}{\hat{\widetilde{\mathbf{R}}}_i}-\frac{\partial h_i}{\hat{\widetilde{\mathbf{R}}}_j}=0$ is also a suitable choice besides $f$. We choose $h$ to be zero in order to reproduce for $Q=0$ the coordinates used in \cite{Sonnleitner2018}. In the main text we used the notation of the usual COM coordinates to refer to the relativistic corrected ones. The only difference is that the vector potential $\Ah(\hat{\mathbf{R}})$ and the scalar Newtonian potential $\phi (\hat{\mathbf{R}})$ should be evaluated at the nonrelativistic COM, but as both change slowly over the size of the atom, this correction is negligible and we can use the relativistic corrected variables instead.

Another major difference arises in the atom-field interaction Hamiltonian~\eqref{HAFmaintext}. In the main text we suppressed a number of terms which are collected in
\begin{align}\label{eq:Hother}
		\hat{H}_{\text{other}}&=
		\frac{1}{8\mu}\left(\hat{\mathbf{d}}\times\hat{\mathbf{B}}(\hat{\mathbf{R}})  \right)^2 
		+\frac{\mu Q^2}{8M^2} \left(\hat{\mathbf{r}}\times \hat{\mathbf{B}}(\hat{\mathbf{R}})  \right)^2\nonumber\\
		&\quad-\frac{m_1-m_2}{4m_1m_2}\left[\hat{\mathbf{p}}\cdot\left(\hat{\mathbf{d}}\times \hat{\mathbf{B}}(\hat{\mathbf{R}})  \right)+\text{H.c.}  \right]\nonumber\\
		&\quad+\frac{Q}{4M}\left[\hat{\mathbf{p}}\cdot \left(\hat{\mathbf{r}}\times \hat{\mathbf{B}}(\hat{\mathbf{R}})  \right) +\text{H.c.}\right]\nonumber\\ &\quad-\frac{Q(m_1-m_2)}{4M^2}\left(\hat{\mathbf{d}}\times \hat{\mathbf{B}}(\hat{\mathbf{R}})  \right)\cdot\left( \hat{\mathbf{r}}\times\hat{\mathbf{B}}(\hat{\mathbf{R}})\right).
\end{align}
The Hamiltonian $\hat{H}_{\text{other}}$ collects all terms which arise from the magnetic field $\hat{\mathbf{B}}$. The $Q$ dependent terms can be interpreted as modifying the electric dipole $\hat{\mathbf{d}}$ by $\hat{\mathbf{d}}'=\hat{\mathbf{d}}+(\mu/M)Q\hat{\mathbf{r}}$ in the limit $m_1 \ll m_2$. They cancel the implicit $Q$-dependence of $\hat{\mathbf{d}}$ such that $\hat{\mathbf{d}}'$ is equal to the dipole moment of a neutral atom (if $m_1 \ll m_2$) \cite{Cormick2011}. Terms contributing also for neutral systems ($Q=0$) are reported and discussed in~\cite{Lembessis1993}.

\section{Time evolution in Ramsey spectroscopy}
\label{appRamsey}

In this appendix we evaluate the time evolution of the internal operators according to the Hamiltonian in Eq.~\eqref{eq:HamRamsey} during the free evolution time, that is, with $\mathbf{E}(\hat{\mathbf{R}},t)=0$. In the Heisenberg picture, the vector of Pauli operators $\hat{\vec{\sigma}}$ evolves as 
\begin{align}
    \frac{d}{dt}\hat{\vec{\sigma}}(t) = \qty(\omega_0+\delta\hat{\omega}(t))\begin{pmatrix}
           0 & 1 & 0\\
           -1 & 0 & 0\\
           0 & 0 & 1
         \end{pmatrix}\hat{\vec{\sigma}}(t)
\end{align}
and the mass defect operator fulfills
\begin{align}\label{eq:massdefectEvolution}
    \frac{d}{dt}\delta \hat{\omega}(t)=-\frac{\ii}{\hbar}\comm*{\hat{H}_{\mathrm{com}}}{\delta\hat{\omega}(t)},
\end{align}
where $\hat{H}_{\mathrm{com}}$ is given in Eq.~\eqref{eq:HcomPaul}. The solution of this equation will be denoted by $\delta \hat{\omega}(t)$ and is independent of the internal DOFs.

We define a time-averaged mass defect operator
\begin{align}
    \overline{\delta\hat{\omega}}=&\frac{1}{T_R}\int_0^{T_R} dt \delta \hat{\omega}(t),
\end{align}
by which the Pauli vector at the end of the Ramsey sequence in first order of the mass defect can be expressed as
\begin{align}\label{eq:internaloperatorsevolution}
    \hat{\vec{\sigma}}(T_R) &= R_z\qty(\omega_0 T_R) \qty[1 +T_R\overline{\delta\hat{\omega}}\begin{pmatrix}
           0 & 1 & 0\\
           -1 & 0 & 0\\
           0 & 0 & 1
         \end{pmatrix}]\hat{\vec{\sigma}}(0) \nonumber\\
     &\simeq R_z\qty\big(\qty(\omega_0+ \overline{\delta\hat{\omega}})T_R)\hat{\vec{\sigma}}(0).
\end{align}
Here $R_z(\theta)$ is the matrix corresponding to a rotation around the $z$ axis by an angle $\theta$. The second line holds to first order in the mass defect and should be understood as shorthand notation for the first line.

The complete Ramsey sequence is given by
\begin{align}
    \hat{\vec{\sigma}}_{\textrm{out}}=&\hat{U}_R^\dagger\qty(0)\hat{U}^\dagger\qty(T_R)\hat{U}_R^\dagger\qty(\omegaL T_R )\hat{\vec{\sigma}}_{\textrm{in}}\hat{U}_R\qty(\omegaL T_R )\hat{U}\qty(T_R)\hat{U}_R\qty(0)\nonumber\\
    =&R_{\vec{n}}\qty(\pi/2)R_z\qty\big(\qty(\omega_0+ \overline{\delta\hat{\omega}})T_R)R_{-y}\qty(\pi/2)\hat{\vec{\sigma}}_{\textrm{in}}\nonumber\\
    =&R_z\qty(\omega_L T_R)R_{-y}\qty(\pi/2)R_z\qty\big(\qty(\omega_0-\omega_L+\overline{\delta\hat{\omega}})T_R)\nonumber\\
    &\times R_{-y}\qty(\pi/2)\hat{\vec{\sigma}}_{\textrm{in}},
\end{align}
where $\vec{n}=-\cos(\omega_LT_R) \hat{\sigma}_y+\sin(\omega_LT_R) \hat{\sigma}_x$. From here it is straight forward to derive Eq.~\eqref{eq:Deltaomega2}. When taking averages, we assume an initial product state of internal and COM DOFs. Moreover, for the stationary COM states considered here, the time average of the mass defect Hamiltonian drops out.

\section{Quantum theory of ion traps}
\label{apptrap}
\subsection{Quasi-Energy-Eigenproblem}

    Here we provide details on the solution to the generalized eigenvalue problem in Eq.~\eqref{eq:eigenvaleq} for the particular case of a Paul trap. Assuming that the potential is separable, it is sufficient to discuss the one-dimensional problem. The treatment follows the one of Glauber~\cite{glauber1992}, which is summarized in \cite{Leibfried2003_review}. The main deviation from these treatments is that we strive to identify $T$-periodic creation and annihilation operators in order to ensure that all states are elements of the enlarged Hilbert space $\mathcal{H}_T$.
    
    We consider the (zeroth-order) Hamiltonian
    \begin{align}
        \hat{H}^{(0)}(t)=\frac{\hat{P}^2}{2m}+\frac{m}{2}W(t)\hat{X}^2,
    \end{align}
    with a real periodic function $W(t+T)=W(t)$. Specifically, we have $W(t)=\mathtt{a}-2\mathtt{q}\cos\qty(\Omega t)$ where $\mathtt{a}$ and $\mathtt{q}$ denote the so-called Mathieu parameters. First, we consider the equation $\ddot{u}(t)=-{W(t)}u(t)$. By the Floquet theorem, solutions can be constructed of the form 
     \begin{align}\label{eq:solu}
     u(t)&=e^{\ii\mu t}v(t),
     \end{align}
     where $0\leq\mu<\Omega$ and $v(t+T)=v(t)$. Following \cite{Leibfried2003_review}, we write this as $\mu=\frac{\beta\Omega}{2}$ and  $v(t)=\sum_{n=-\infty}^\infty C_{2n}\e^{\ii n\Omega t}$. In order to relate with the main text, we have to set $\mu=\omega_i$. The real-valued coefficients $\beta$ and $C_{2n}$ can be determined as in \cite{Leibfried2003_review}. 
     
     We also adopt the normalization condition $u(0)=\sum_n C_{2n}=1$, which implies for the time derivative $\dot{u}(0)=\ii\nu$, $\nu=\Omega\sum_nC_{2n}\qty(\beta/2+n)$. Here $u^*(t)$ is also a linearly independent solution, which is likewise assumed to be normalized $u^*(0)=1$ so that $\dot{u}^*(0)=-\ii\nu$. The Wronskian $w=u(t)\dot{u}^*(t)-u^*(t)\dot{u}(t)$ is time independent, since $\dot{w}=u(t)\ddot{u}^*(t)-u^*(t)\ddot{u}(t)=-W(t)u(t){u}^*(t)+W(t)u^*(t){u}(t)=0$. 
    Thus, $w=u(0)\dot{u}^*(0)-u^*(0)\dot{u}(0)$ is fixed by the initial conditions for the linearly independent solutions in $u(t)$ and $u^*(t)$. For the specific choice made above, we thus have for all times $u(t)\dot{u}^*(t)-u^*(t)\dot{u}(t)=-2\ii\nu$. In lowest order of the Mathieu parameters we find (cf. \cite{Leibfried2003_review})
    \begin{align*}
        \beta&\simeq \sqrt{\mathtt{a}+\frac{\mathtt{q}^2}{2}}, &
        v(t)&\simeq \frac{1+\frac{\mathtt{q}}{2}\cos(\Omega t)}{1+\frac{\mathtt{q}}{2}}, &
        \nu&\simeq \mu.
    \end{align*}
    
    Based on this solution, we define the explicitly time-dependent operator (in the Schr\"odinger picture)
    \begin{align}\label{eq:anop}
        \hat{a}(t)=\frac{\ii\e^{-\ii\mu t}}{\sqrt{2\hbar m\nu}}\qty(u(t) \hat{P}-m \dot{u}(t) \hat{X}),
    \end{align}
    which effectively depends on the functions $v(t)$ and $\dot{v}(t)$ only and therefore is by construction periodic in time with period $T$. Here $\nu$ is a frequency which will be chosen later. Note that for $t=0$ this corresponds to the expression for the annihilation operator for a harmonic oscillator with frequency $\nu$, that is, $\hat{a}(0)=\sqrt{\frac{m\nu}{2\hbar}} \hat{X}+\ii \frac{1}{\sqrt{2\hbar m\nu}}\hat{P}$.
    
    Due to the constant value of the Wronskian, $\hat{a}(t)$ and its adjoint operator $\hat{a}^\dagger(t)$ satisfy bosonic commutation relations at equal times $\qty[\hat{a}(t),\hat{a}^\dagger(t)]=1$. The operator $\hat{a}(t)$ has a unique (up to a global phase) eigenstate $\ket{0,t}$ of eigenvalue $0$,  $\hat{a}(t)\ket{0,t}=0$, which can be constructed by projecting this equation into the position representation $(\ii\hbar u(t)\partial_x+m\dot{u}(t) x)\braket{x}{0,t}=0$, giving the normalized solution 
    \begin{align}
        &\braket{x}{0,t}=\qty(\frac{m\nu}{\pi\hbar})^{1/4}\frac{\e^{\ii\mu t/2}}{u(t)^{1/2}}
        \exp\qty[\frac{\ii m}{2\hbar}\frac{\dot{u}(t)}{u(t)}x^2]\label{eq:wavefunc}\nonumber\\
        &=\qty(\frac{m\nu}{\pi\hbar})^{1/4}\frac{1}{v(t)^{1/2}}
        \exp\qty[-\frac{\mu m}{2\hbar}\qty(1-\frac{\ii }{\mu}\frac{\dot{v}(t)}{v(t)})x^2].
    \end{align}
    The global phase in \eqref{eq:wavefunc} is chosen so as to make the state $\ket{0,t}$ periodic with period $T$ and have $\ket{0,t}\in\mathcal{T}\otimes\mathcal{H}$. The periodicity is evident in the second line. Using Eq.~\eqref{eq:wavefunc}, it can be shown by direct calculation that
    \begin{align}
        \qty(\hat{H}^{(0)}(t)-\ii\hbar\partial_t)\ket{0,t}&=E_0\ket{0,t}, & E_0&=\frac{\hbar\mu}{2}.
    \end{align}
    
    We easily verify that the creation and annihaliton operators satisfy eigenoperator equations with respect to the generalized Hamiltonian
     \begin{subequations}\label{eq:bosonicCOMs}
    \begin{align}
        \qty[\hat{H}^{(0)}(t)-\ii\hbar\partial_t,\hat{a}(t)]&=-\hbar\mu a(t),\\
        \qty[\hat{H}^{(0)}(t)-\ii\hbar\partial_t,\hat{a}^\dagger(t)]&=\hbar\mu a^\dagger(t).
    \end{align}
    \end{subequations}
    The bosonic commutation relation and the commutators in \eqref{eq:bosonicCOMs} are identical to those of a time-independent harmonic oscillator. Thus, the same algebra used there can be applied here to show that
    \begin{align*}
        &\qty(\hat{H}^{(0)}(t)-\ii\hbar\partial_t)\ket{n,t}=E_n\ket{n,t},\\
        &E_n=\hbar\mu\qty(n+\frac{1}{2}),\quad \ket{n,t}=\frac{1}{\sqrt{n!}}\qty(a^\dagger(t))^n\ket{0,t}.
    \end{align*}
    Here it is crucial to have constructed the annihilation operators to be $T$ periodic, as otherwise the Fock states would not be proper elements of $\mathcal{H}_T$. Note that the quasienergy eigenvalues should respect $E_n<\hbar\Omega$, which will be violated for some $n$. However, for the physically relevant case where $\mu\ll\Omega$, that is, $\beta\ll 1$, this is of no concern practically.

\subsection{Ion trap with linear potential}

In order to study both the cases of dc forces and gravity, we will consider now a trap potential with a linear potential $-F\hat{X}$ such that the generalized Hamiltonian is
    \begin{align}
        \hat{H}(t)&=\hat{H}^{(0)}-F\hat{X}-\ii\hbar\partial_t.
    \end{align}
    We seek operators of the form
    \begin{align}
        \hat{b}(t)=\hat{a}(t)+\alpha(t),\label{eq:abar}
    \end{align}
    which fulfill the eigenoperator equation $[\hat{H}(t),\hat{b}(t)]=-\hbar\bar{\mu}\hat{b}(t)$, with a $T$-periodic function $\alpha(t)$ and a new eigenfrequency $\bar{\mu}$ to be determined. In order for $\alpha(t)$ not to be an operator, we find that $\bar{\mu}=\mu$. Therefore, the differential equation that $\alpha(t)$ needs to obey is
    \begin{align}
        \dot{\alpha}(t)
        +\ii{\mu}\alpha(t)=-\frac{\ii Fe^{-\ii\mu t}u(t)}{\sqrt{2\hbar m \nu}}
        =-\frac{\ii F v(t)}{\sqrt{2\hbar m \nu}}.
    \end{align}
    Imposing $T$ periodicity on $\alpha\left(t\right)$, we find in lowest order of the Mathieu parameters
    \begin{align}
        \alpha(t)=&-\frac{F}{\sqrt{2\hbar m \nu}}\left[\frac{\mu}{\mu^2-\Omega^2}\e^{-\ii\mu t}u(t)\right.\nonumber\\
        &\left.-\frac{1}{1+\frac{\mathtt{q}}{2}}\frac{1}{\mu\left(\mu^2-\Omega^2\right)}\left(\Omega^2+\ii\frac{\mathtt{q}}{2}\mu\Omega\sin\left(\Omega t\right)\right)\right].\label{eq:alpha}
    \end{align}
    With this solution we have determined the eigenoperator of the trap potential including a linear force. In the main text, this formalism is applied for each Cartesian direction $i$ where we need to use the replacement $F\rightarrow F_i$ and $\mu\rightarrow \omega_i=\beta_i\Omega/2$. In the case of gravity and a spurious $E_{dc}$ field this will correspond to $F_i=Q\mathrm{E}_{\mathrm{dc},i}-Mg_i$. This will lead to a total fractional frequency shift of
    
\begin{align}
    \llangle[\Big]{\frac{\delta\hat{\nu}}{\nu_0}}\rrangle[\Big]={}& \llangle[\Big]{\frac{\delta\hat{\nu}}{\nu_0}}\rrangle[\Big]_{\!\mathbf{n}}
    +\sum_{i=1}^3\frac{8g_i\left(QE_{\text{dc},i} - Mg_i\right)}{M\Omega^2c^2\left(2\mathtt{a}_i +\mathtt{q}_i^2\right)}\nonumber\\
    &-\left(\frac{2 \left(QE_{\text{dc},i}  - Mg_i\right) \mathtt{q}_i}{M c \left(2\mathtt{a}_i +\mathtt{q}_i^2\right)\Omega}\right)^2.
\end{align}

In evaluating this and similar expressions in the main text it is necessary to deal with fractions of polynomials in $\mathtt{a}$ and $\mathtt{q}$. In order to simplify these expressions, we associated a small parameter $\epsilon$ via the substitution $\mathtt{a}\rightarrow \epsilon^2 \mathtt{a}$ and $\mathtt{q}\rightarrow \epsilon \mathtt{q}$, assuming $\mathtt{q}^2, \abs{\mathtt{a}} \ll 1$.  Finally, we performed a Taylor expansion of the rational function in terms of $\epsilon$ and maintained the relevant contributions.

\end{document}